Michael Bassi[1], Giovanni Salucci[2]

# Guidelines for the Enhancement of the Corpus and the Verismo Vocabulary

**Abstract**: *VIVer is a digital lexicography project with historical-literary and historical-linguistic aims that can be considered a case study of a Digital Humanities project. This paper presents the IT choices made to promote the dissemination and enhancement of the results, analysing the issues and advantages for wider adoption, beyond the specific VIVer project, serving as a model and inspiration for future projects.*

## 1. Introduction. The reference scenario

VIVer is a digital lexicography project aimed at providing an organic description of Italian Verismo. It consists of the corpus – composed of texts belonging to the textual heritage of Italian Verismo authors, selected by the scientific coordinators of the project for their representativeness of Verismo Italian – and the dynamic electronic vocabulary, innovative for its structure and consultability (**Alfieri et al. 2024**).

For Digital Humanities (DH) projects in the linguistic and literary field, such as VIVer, there are well-established guidelines and methodological approaches for the design and coordination phase, for the adoption of standards, for digitization procedures – from text transcription to linguistic analysis – and for adopting open-access publishing policies for results (see for example **Schwandt 2021**); in general reference is made to the so-called FAIR principles.

### 1.1 The FAIR principles

---

[1] Università degli studi di Firenze, ORCID: 0009-0008-0569-3703
[2] Università degli studi di Firenze e Accademia della Crusca, ORCID: 0000-0002-9587-1620

Formalized in 2016 (**Wilkinson et al. 2016**), the FAIR guidelines lay the foundations for research data management in any field. Specifically, the principles refer to three types of entities: data (or any digital object), metadata (information about that digital object), and the digital infrastructure. For example, one of the principles contained in the guidelines states that both metadata and data should be registered or indexed in a searchable resource (the infrastructure component).

From the initials of the acronym FAIR it is possible to identify the cornerstones of these good practices:

F - FINDABLE. The first step in (re)using data is to find them. Metadata and data should be easy to locate both for users and for computational procedures – such as aggregators, search engines, or AI agents. Like data, metadata must be in a machine-readable format, essential for the automatic detection of datasets and services.

A - ACCESSIBLE. Once the user (or machine) finds the data, it is necessary to know its access policies, including any authentication or authorization.

I - INTEROPERABLE. Data are usually integrated with other data, forming an increasingly broad and complex information system. Therefore, data (and metadata) must interact with applications or workflows for analysis, archiving, and processing.

R - REUSABLE. The main objective of FAIR is to optimize data reuse. To achieve this, data (and metadata) must be well described so that they can be replicated and/or combined in different contexts.

These principles underpin the main guidelines and mandates that accompany scientific research funding; for example, the Horizon Europe call[3] explicitly cites the FAIR principles and links them to the obligation to publish results in open access, depositing a copy (accepted version or final publisher version) in a trusted repository, with the most open possible Creative Commons licenses. The document clarifies the concept of a trusted repository, specifying its characteristics (**Horizon Europe 2024**).

In Italy too, research data management is a crucial issue, and the main calls for funding from the MUR or other institutions provide specific guidelines for adopting Open Science policies and for managing data according to FAIR principles, to ensure greater transparency and accessibility of research, in line with European and international best practices (**Delle Donne 2025**).

**1.2 Eligible outputs for scientific evaluation**

---

[3] https://horizoneurope.apre.it/

Results produced in DH projects are often presented as databases, websites, blogs, podcasts, or digital publications in various forms; for authors it is therefore difficult to obtain recognition of credit and to ensure traceability.

In the Call for the Evaluation of Research Quality (**VQR 2023**), which regulates the methods for evaluating institutions that produce scientific research in Italy, Article 5 lists the research products eligible for evaluation; in addition to typical publication formats (monographs, edited volumes, journal articles, chapters in collective volumes or contributions in conference proceedings), point (e) introduces other types of publications (including those typical of DH projects); however, these products will be eligible only if accompanied by official elements enabling the identification of the author and the date of production.

This requirement can be supported by the systematic adoption of the DOI (Digital Object Identifier), which is an interactive system for persistent identification and interoperable exchange of information on the web (**Paskin 2005**). In the metadata for DOI registration are contained the information about the object's identification, dating, and authorship traceability; it is also possible to use portions of metadata provided in the registration schema for another purpose, far from secondary, namely to favor the dissemination of the resource to which the DOI refers, through a wide set of descriptive metadata, in accordance with FAIR principles. In particular, referring to lexicographic projects, the DOI represents the ideal tool for the considerations we will develop below, being, as already mentioned, a system of persistent identification of digital, physical, or logical objects widely used in academic publishing and in the archiving of scientific experiment datasets, but scarcely used so far in the humanities (**Salucci 2023**). Precisely the systematic adoption of DOI with a rich set of metadata constitutes one of the new guidelines proposed here.

## 2. Strategies for dissemination and promotion through the application of the FAIR principles

The VIVer project provides for the systematic use of descriptive formats in XML language, which allow easier data exchange, and for the adoption of appropriate procedures and guidelines (Open Standards) that guarantee an adequate level of quality of the results obtained, certifying their scientific value in line with generally adopted policies. The use of standard protocols also enables the dissemination and retrieval of information on the Internet by archives and digital libraries, while respecting privacy protection and the intellectual property rights of rights holders.

### 2.1 Prerequisites: the platform and the staff

One critical aspect to consider is that using metadata and DOI for the enhancement of DH projects nonetheless has an entry barrier, consisting of financial aspects but above all technical and scientific ones, since it requires an adequate IT infrastructure and specific skills for the management of high-quality metadata, in addition to the ability to adopt an appropriate policy of digital archiving (**Salucci 2022**).

The platform chosen for the Verismo project (described in **Alfieri et al. 2024**) is **WCM digital humanities**[4], developed by *Progettinrete* for the enhancement of research outputs in the DH sector. The WCM platform, through various operational environments and modules, provides digitization services, advanced management and markup of research products and tools (dictionaries, glossaries, archival collections, corpora of various kinds), management of metadata and DOI, implementation of digital preservation policies, integration of search engines, and monitoring of statistics to measure impact.

With this IT infrastructure, the core working group responsible for the design has established the guidelines and operational methods to implement them, subsequently involving, to varying degrees, all participants who are involved in the project at different stages of its development and completion.

The scope of the entire operation is clearly characterized by its predominantly IT nature, but it also involves the indispensable participation of those who manage the data and metadata throughout the project's implementation process.

**2.2 Metadata for dissemination**

To guarantee the results that we will now describe in detail, it will be necessary to systematically adopt persistent identifiers such as ORCID, ISBN, ISSN, DOI, ISNI, and so on. It will be precisely through the relationships that will be established between these identifiers, recorded in the metadata accompanying research outputs, that an informational network will be built, enabling navigation from entity to entity and promoting dissemination (**Macgregor et al. 2023**).

The creation of high-quality metadata will be a significant aspect throughout the project, both in the initial choice of profiles and tags to be used, and in the subsequent care to be given to the metadata activity object by object; metadata are in fact useful not only to classify and manage digital objects within the collection and repository, but above all in the perspective of interoperability with other systems (see the Digital Library[5] being developed under the PNRR). In this way, data and metadata will be distributed under appropriate licenses on the portal and on other online aggregators.

To describe this decidedly innovative aspect, it seems useful to explore the choices made and their motivations. Considering the specificities of the project, which, in addition to the scientific and lexicographic dimension, aims at reaching a broader audience, the choice of metadata adopted both specialized and general profiles. First of all, we planned the use of **Dublin Core**[6]; in addition to managing the **CrossRef profile**[7] (for DOI registration according to the dataset schema[8]); we also opted to adopt **Highwire Press**[9], to facilitate indexing in Google Scholar and academic aggregators. Finally, to optimize dissemination and sharing on social networks—which should not be underestimated, especially when one aims to address a generalist audience—we decided to adopt **Open Graph**[10] and **Twitter Cards**[11] profiles. Similar choices have already been made in analogous projects, and they have had an immediate and positive impact on indexing results in Google and other search engines, both general and academic: only a few days after the public release, the lexicographic records and digital editions were indexed correctly and appeared among the top search results for relevant keywords.

---

[4] https://www.wcm.it/wcm-digital-humanities/
[5] https://digitallibrary.cultura.gov.it/
[6] https://www.dublincore.org/specifications/dublin-core/dces/

There will also be procedures and methods for measuring the impact and results of dissemination, including trends and absolute numbers regarding access, views, and downloads of materials.

**2.3 Use of DOI and metadata for registration**

VIVer is configured as a digital lexicographic workshop in which scientific activities are strongly integrated with IT ones. In digital lexicographic projects, the use of persistent identifiers is recommended, and thus, as already mentioned earlier, it was decided to use the DOI (Digital Object Identifier) as a tool for dissemination and promotion, choosing **CrossRef** as the registration agency (**Salucci 2022**). The choice of CrossRef is strategic for the completeness of the schema provided by the registration system, and for the interconnections among permanent identifiers that are established even automatically (for example, among bibliographic reviews); furthermore, by using the **Crossmark service**[12], it will be possible to track the different versions of materials, as will be discussed later.

The first step is to assign and register a DOI for each digital edition of new works in the corpus, also digital (produced in XML, HTML, and PDF formats). Being new editions in every respect, DOI registration makes it possible to certify and record the responsibility of each researcher, documenting their role in the metadata themselves.

But the real innovation lies in the assignment and registration of a series of DOIs also for the vocabulary database, corresponding to the various hierarchical levels with which the VIVer database is organized (database, collection, record — the latter corresponding to the individual vocabulary entry).

**2.4 Long-term archiving**

---

[7] https://www.crossref.org/documentation/schema-library/metadata-deposit-schema-5-3-1/
[8] https://www.crossref.org/documentation/research-nexus/datasets/
[9] https://scholar.google.com.au/intl/en/scholar/inclusion.html
[10] https://ogp.me/
[11] https://developer.x.com/en/docs/x-for-websites/cards/overview/markup
[12] https://www.crossref.org/services/crossmark/

One of the problems that most afflicts scientific publishing, and in particular native digital Open Science, is the lack of attention generally paid to long-term preservation archiving. If it is already difficult and expensive to keep native digital resources operational, due to the rapid obsolescence of IT systems and their continuous and, in some respects, sudden innovation, it is even more difficult to organize long-term policies; all the more so in a context where even academic digital publications (journal articles and monographs in PDF format, much easier to manage) largely lack safeguarding and archiving plans (**Eve 2025**).

The decision to implement a long-term preservation plan for the entire project from the outset, as described below, is clearly an innovative aspect.

**2.5 Versioning with responsibility through DOI metadata**

Another particularly relevant and widespread issue, especially when dealing with databases (for example, lexicographic ones like VIVer), is that of citability: not only because data usually have non-homogeneous granularity, but also, and above all, because of the lack of versioning. Since these databases are open and updatable, the information they contain is always potentially modifiable (both in the publication address and in the content itself). Moreover, even if citation systems were present, there is no guarantee of reproducibility and durability of the cited content over time (**Bach et al. 2022**).

The current situation is so precarious that more than once one is forced to retrieve, when possible, the cited online resource—made modified or unavailable—through services such as Wayback Machine[13], whose incompleteness or unreliability is well known. It may be frightening, but this is the situation in which certain types of digital data remain: a sort of periodic and sudden Fahrenheit 451 where books are not burned but deleted or modified, without any warning, without any countermeasures possible at that point. To overcome this problem, of course, it is possible to introduce persistent identifiers such as DOIs, but this alone is not enough; it is in fact necessary to pair the systematic use of persistent identifiers with good practices of content preservation, as required (but unfortunately not always implemented) in academic publishing.

It is necessary to introduce into the flow of database publications the same policies that accompany, for example, scientific articles published in journals; except for corrections of mere typos or material errors, in all other cases where changes are made to content after its publication, it is necessary to produce a new version and modify the DOI metadata through the Crossmark service. In this way, the user will be informed of changes and can eventually access the previous version.

All the more so, in the case of substantial changes (such as the intervention of a new author), it is necessary to preserve the previous version and publish a new one, registering a new DOI; in the registration metadata, care will be taken to link it to the previous one, to track changes as well as provenance; on the record (and in its PDF version) the specific

---

[13] https://web.archive.org/

Crossmark button will be inserted to access the version history and verify the currency of what is being consulted.

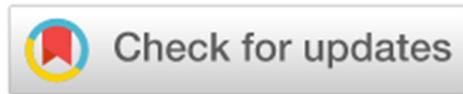

Image 1 – Crossmark button that allows you to check the version history of the displayed content.

## 3. Application of the guidelines to the VIVer project

### 3.1 New digital editions with ISBN and DOI for the works in the corpus and the Verismo Vocabulary

The XML-TEI version of texts represents, in every respect, a kind of edition (even critical) of the original text, produced by researchers. We should not be misled by the different syntactic rules: the linguistic phenomena highlighted through the application of XML tags have the same informative value as what could be conveyed by a canonical description in any modern language; in this sense, a critical edition with text and notes can be entirely equated to a text marked with specific tags (to annotate, for example, lemmas, authors, works, personalities, toponyms, etc.); indeed, in this second case, the presence of a defined and standardized descriptive structure in XML-TEI guarantees an easier reading and extraction of relevant information by automated computer systems.

It is therefore natural to apply to these new editions the same process of scientific validation and publication that has so far been reserved for canonical scholarly editions. By using appropriate scholarly series in the catalog of academic publishers, these studies can be subjected to peer review and published, assigning them a publisher DOI and giving an e-ISBN to each distributed digital format (XML, HTML, EPUB).

### 3.2 Citability and versioning of the Verismo Vocabulary

The Verismo Vocabulary, which is currently under development, represents an emblematic and real case of an open, integrable, updatable database; the degree of expansion is not at present predictable because the project is still rapidly evolving, with the participation of additional research units involving new scholars.

The database expansion will occur both in the Corpus and in the Vocabulary; for the latter, over time, not only the inclusion of new entries is expected, but also the revision and modification of those already published, in order to take into account the results emerging from new works added to the corpus and from the intervention of new researchers involved in varying degrees in the revision of what has already been published online.

To facilitate the revision of lexicographic entries — for example, retrieving previous choices made by the author of an entry (to add new contexts where relevant or to modify the date of first attestation, etc.) — it has been decided to introduce a field indicating the date of the last check (and indexing) of the corpus; in this way, researchers are given the possibility to carry out a diachronic study of the corpus texts managed through the platform, and it becomes possible to track the data history.

In publishing the entries, it was therefore decided to use versioning: the DOI with its related metadata is recorded at the time of the entry's first online publication; in case of modifications, the system automatically creates a new version of the entry and archives the previous one, leaving it to the responsible editor to decide whether to keep the existing DOI or register a new one. The previous version of the entry thus remains available on the portal; the link between the two versions of the entry is recorded in the metadata, in the part of the schema reserved for Crossmark, and made available within the entries themselves (or in the generated PDF) through the Crossmark "Check for updates" service.

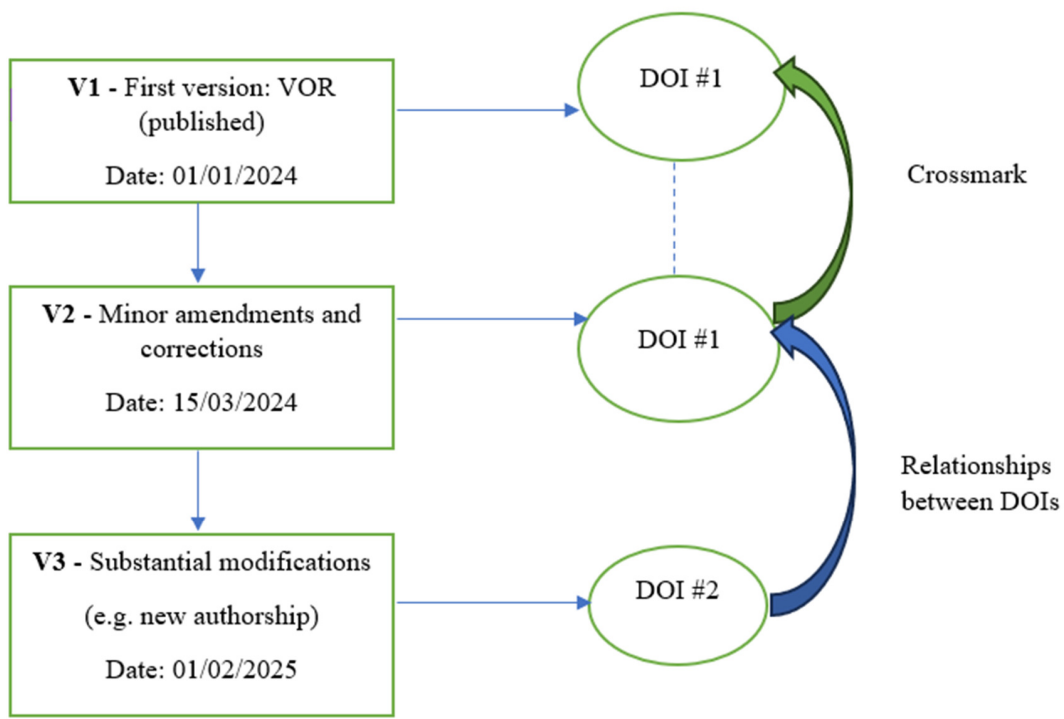

Image 2 – Diagram showing the correspondence between the different versions of the same record and their respective DOIs.

**3.3 Repository with long-term archiving of the Verismo corpus**

In the VIVer project, the creation of the "Reading Room" is planned, which represents and collects the corpus texts in various formats. For each text, there may be a collection of images for facsimile consultation, digitization in PDF (with or without OCR), the text revised and corrected by the editorial staff, and the text marked in XML-TEI. Each format will be accompanied by specific metadata containing information about the reliability of the text and the processing procedures, in addition to the usual information on copyright and reuse licenses. Such a plurality of formats will make the entire project not only user-friendly but also machine-friendly, allowing its propagation on the web and consultability at various levels.

It has therefore been decided to define and adopt a policy of long-term archiving and preservation of materials, using the functions present in the **WCM Repository** module, appropriately configured.

In this way, the digital materials to which the policies are to be applied are cataloged, in the most common formats (images, texts, apparatus, etc.), all accompanied by detailed records drafted in compliance with international scholarly standards. The policy provides for cloud archiving on resources managed by the **Accademia della Crusca**[14] and parallel archiving on a reliable repository: the chosen system is **Zenodo**[15], the most important repository of Open Science, managed by CERN and supported by the European Community, which has chosen it as the official repository of Open Science. This guarantees the support of a large, reliable, and secure IT infrastructure.

The policy defined for VIVer is already prepared to handle any malfunctions of the IT infrastructure: the use of the DOI makes it possible to quickly modify the URL for accessing resources, keeping them available; in ordinary circumstances, the pointing is to the Crusca archiving system, but this can be updated if necessary by replacing it with Zenodo, thus activating the backup and disaster recovery policy through the long-term preservation environment.

### 4. Conclusions

VIVer can be considered a "hybrid" digital lexicography project, straddling the electronic dictionary and the computerized literary archive, and in this sense embodies a typical DH project in the linguistic and literary field; it can therefore serve as a case study for other institutions interested in the digital enhancement and communication of lexicographic and literary heritage.

In its conformity to similar projects, thanks to the adoption of the guidelines described above, this specific humanities-digital project is also a pioneer in addressing issues that are both much-debated and deeply rooted within the sector.

---

[14] https://accademiadellacrusca.it/
[15] https://zenodo.org/

The strategic operational choices adopted for the promotion and dissemination of VIVer are certainly innovative but carried out in profound adherence to the FAIR principles; they have been described in this contribution to encourage their adoption as guidelines for similar projects, in the spirit of collaboration and sharing of scholarly expertise. The use of international standards and of established platforms and services (Zenodo, DOI, Crossref, etc.) now represents an indispensable condition for implementing good, sustainable, and replicable sharing practices.

## Acknowledgments
The authors thank President Gabriella Alfieri and the entire Scientific Committee of the VIVer project for their availability and attention in welcoming the innovations and contributing to their implementation; as well as Professor Marco Biffi, who, in his role as IT manager of the project, strongly wanted to embark on this path.

## Conflict of interest
Author Giovanni Salucci reports a potential conflict of interest as he serves as the CEO of the company that developed the WCM software used in the VIVer project.